\documentclass[a4paper, 11pt]{article}
\usepackage[pdftex, a4paper]{geometry}
\usepackage{graphicx}				

\usepackage{amsmath}
\usepackage{amsfonts}
\usepackage{dsfont}
\usepackage{mathrsfs}
\usepackage{amssymb}
\usepackage{bbm}
\usepackage{epsfig}
\usepackage[english]{babel}
\usepackage[all]{xy}
\usepackage{color}

\newcommand*\mcapinn[2]{\vcenter{\hbox{$\mathsurround=0pt
  \ifx\displaystyle#1\textstyle\else#1\fi\bigcap$}}}

\newcommand*\mcupinn[2]{\vcenter{\hbox{$\mathsurround=0pt
  \ifx\displaystyle#1\textstyle\else#1\fi\bigcup$}}}

\DeclareFontFamily{OT1}{pzc}{}
\DeclareFontShape{OT1}{pzc}{m}{it}{<-> s * [1.200] pzcmi7t}{}
\DeclareMathAlphabet{\mathpzc}{OT1}{pzc}{m}{it}

\newtheorem{theorem}{Theorem}

\begin{document}

\title{\bf Reaching Agreement  in Quantum Hybrid Networks}
\author{ Guodong Shi, Bo Li, Zibo Miao, Peter M. Dower, and Matthew R. James\thanks{G. Shi and M. R. James are with the Research School of Engineering, The Australian National University, ACT 0200, Canberra, Australia.    Email:  guodong.shi, matthew.james@anu.edu.au}
\thanks{B. Li is with Key Lab of Mathematics Mechanization,  Chinese Academy of Sciences, Beijing 100190,
China. Email: libo@amss.ac.cn}
\thanks{ Z. Miao and P. M. Dower are with Department of Electrical \&
Electronic Engineering, The University of Melbourne, Parkville, Victoria
3010, Australia. Email: zmiao, pdower@unimelb.edu.au}}
\date{}
\maketitle
\begin{abstract}
 We consider a basic quantum hybrid  network model consisting of  a number  of nodes each holding a qubit, for which the aim is to drive the network to a consensus in the sense that all qubits  reach a common state. Projective measurements are applied serving as control means, and the measurement results  are exchanged  among the nodes  via classical communication channels.  We show how to carry out centralized optimal path planning for this network with all-to-all classical communications, in which case the problem becomes a stochastic optimal control problem with a continuous action space. To overcome the computation and communication obstacles facing  the centralized solutions, we also develop a distributed  Pairwise Qubit Projection (PQP) algorithm, where  pairs of nodes meet at a given time and respectively perform  measurements at their geometric average. We show that the qubit states are driven to a consensus almost surely  along the proposed PQP algorithm, and that the expected qubit density operators converge to the  average of the network's initial values.
\end{abstract}

\section{Introduction}
Consensus seeking over complex networks has played a foundational role in the development of distributed computation  and networked control systems \cite{lynch,magnus}.  How a set of isolated processors  communicating only by means of two-party messages  reach  a common state in the presence of faulty nodes was a prior concern for fault-tolerant distributed computation \cite{Pease:1980:RAP:322186.322188}. Distributed controller design that drives a network of autonomous agents to certain  consensus state such as the network average or some leader's state \cite{Jadbabaie2003} turned  out to be a primary step towards control, estimation, and optimization of networked control systems \cite{magnus}. In the past decades, tremendous research efforts have been devoted to efficient design and convergence analysis of consensus and synchronization  algorithms  motivated by various social, engineering, and physical systems, e.g., \cite{Boyd2006,Boyd-SIAM-2006,Kar-TSP-2010,Vicsek1995,Golub-naivelearning-2010}.

In particular, consensus over quantum networks where node states are in quantum space and algorithms must be implemented by feasible quantum means has drawn attention \cite{Ticozzi,Shi-TAC}. Quantum particles (subsystems) can be interconnected by local environments which are by themselves also quantum systems,   the resulting state evolution will lead to a symmetric state consensus over such a quantum network, a concept introduced in \cite{Ticozzi}. The reduced states of the nodes will in turn asymptotically tend to the average  of the nodes' initial reduced states, in the almost sure sense along the discrete algorithm proposed in \cite{Ticozzi} and deterministically along the master equation approach proposed in \cite{Shi-TAC}. Such methods are essentially coherent quantum control for open quantum systems \cite{rivas2011open},  where the  involved  local environments
can only be engineered at a small scale. On the other hand,   many types of quantum networks, especially quantum communication networks, are hybrid in the sense that both quantum and classical parts co-exist \cite{cirac2007,PhysRevLett.103.240503}. Quantum operations (often being measurements) can be performed locally and then the outcomes of the measurements are exchanged via classical communications, leading to the so-called local-operation classical-communication (LOCC) networks which have served  as  protocols for quantum cryptography or potential tools for engineering complex quantum states \cite{CiracRandomNetworks2011}. Measurement-based quantum control has also been demonstrated  as effective means of manipulating quantum states both theoretically and experimentally   \cite{Rabitz-PRA,wiseman2011quantum,Shi-PRA14,QubitFeedback2014}.

In this paper,  we consider a consensus seeking problem over a quantum hybrid  network  consisting of  a number  of nodes each holding a qubit, where projective measurements are applied  and the measurement results  are exchanged.  The problem of centralized optimal path planning for the network with all-to-all classical communications is shown to be a stochastic optimal control problem, whose computation and communication complexities are analyzed. We also develop a distributed  Pairwise Qubit Projection (PQP) algorithm, where  pairs of nodes meet at a given time and respectively perform  measurements at their geometric average. The qubit states are driven to a consensus almost surely along the proposed PQP algorithm. The expected qubit density operators actually  converge to the  average of the network's initial values, consistent with the work of \cite{Ticozzi,Shi-TAC} for open quantum networks.

The remainder of this paper is organized as follows. Section \ref{sec2} presents some basic  preliminaries on quantum states and measurements,  and then introduces the considered hybrid quantum network model. Section \ref{sec3} and Section \ref{sec4} investigate centralized and distributed solutions to the considered qubit agreement problem, respectively. Finally a few concluding remarks are given in Section \ref{sec5}.

\section{Preliminaries and The Model}\label{sec2}
In this section, we first present some preliminaries on quantum states and quantum measurements \cite{Nielsen}, and then introduce the hybrid quantum network model under investigation.
\subsection{Quantum States and Measurements}
  The state space associated with any isolated quantum system is a complex vector space with inner product, {i.e.}, a Hilbert space $\mathcal{H}$. The system is completely described by its state vector, which is a unit vector in the system's state space and often denoted by $|\psi\rangle\in\mathcal{H}$  known as the Dirac notation. For an open quantum system, its state can also be described by a positive (i.e., positive semi-definite) Hermitian density operator  $\rho$ satisfying $\text{tr}(\rho)=1$. Let $\big(\cdot,\cdot\big)$ be the inner product equipped by the Hilbert space $\mathcal{H}$. Under Dirac notion this inner product is written as $\big(|\psi \rangle, |x\rangle\big)=\langle\psi |x\rangle$, where $\langle\psi |$ is the dual vector of $|\psi \rangle$.  A quantum state $|\psi\rangle\in\mathcal{H}$, induces a density operator, namely  $\rho=|\psi\rangle\langle\psi |$ by
\begin{align*}
  |\psi\rangle\langle\psi | \Big(|x\rangle\Big)=  \Big( \langle\psi |x\rangle \Big)  |\psi\rangle,\  |x\rangle \in \mathcal{H}.
  \end{align*}
 Density operators provide a convenient description of {\it mixed states} as ensembles of pure states: If a quantum system is in state $|\psi_i \rangle$ with probability $p_i$ where $\sum_i p_i=1$, its density operator is
  \begin{align*}
  \rho=\sum_i p_i   |\psi_i\rangle\langle\psi_i |.
  \end{align*}
Any  positive and Hermitian operator  with trace one defines a proper density operator describing certain quantum state, and vice versa.

 A projective measurement is described by an {\it observable} being  a Hermitian operator over the state space $\mathcal{H}$ of the system being observed. Let the dimension of $\mathcal{H}$ be $m$ and let $O$ be an observable with nondegenerate spectrum, i.e.,   the eigenvalues  $O_1,\dots, O_m$ of  $O$ are  distinct. Let $|1\rangle,\dots, |m\rangle$ be the eigenvectors  corresponding to eigenvalues $O_1,\dots, O_m$ of $O$, respectively. As $O$ is Hermitian,   $|1\rangle,\dots, |m\rangle$  form a complete basis of the Hilbert space $\mathcal{H}$. Consequently,  $| \psi \rangle$ can be written as
$$
| \psi \rangle= \sum_{j=1}^m c_j |j\rangle,
$$
where the $c_j$ are complex numbers satisfying $\sum_{j=1}^m |c_j|^2 =1$. Let a system  be prepared in state $| \psi \rangle$ where the measurement $O$ is performed. Then the outcome of such measurement is random taking value in    $\{O_1,\dots, O_m\}$, for which the probability of observe outcome $O_j$ is
$$
\mathbb{P}(O_j)=|c_j|^2.
$$
Moreover, the state after the measurement becomes  $|j\rangle$ if $O_j$ is observed.

\subsection{A Hybrid Quantum Network Model}

Let a network of nodes be indexed in the set $\mathrm{V}=\{1,\dots,N\}$. Each node holds a qubit, i.e., a quantum system whose state space $\mathcal{H}$ is a two-dimensional Hilbert space.  Let $|0\rangle$ and $|1\rangle$ form an orthogonal basis of the qubit space $\mathcal{H}$. Projective measurements can be performed at the individual qubits, respectively.
An available  projective measurement $\mathsf{M}_\alpha$ is described by its two eigenstates
\begin{align*}
\cos \alpha|0\rangle +\sin \alpha |1\rangle,
\end{align*}
and
\begin{align*}
\cos\big(\alpha+\pi/2\big)|0\rangle +\sin \big(\alpha+\pi/2\big) |1\rangle.
\end{align*}
We assume that the measurements are in the set
$$
\mathpzc{M}=\Big\{\mathsf{M}_\alpha: \alpha \in[0,\pi/2)\Big\}.
$$
The outcomes of a measurement $\mathsf{M}_\alpha$ are indexed by $\ltimes$,  corresponding to eigenstate $\cos \alpha|0\rangle +\sin \alpha |1\rangle$, and $\rtimes$,  corresponding to eigenstate $\cos(\alpha+\pi/2)|0\rangle +\sin (\alpha+\pi/2) |1\rangle$. For the ease of presentation we will sometimes  identify a measurement in the set $\mathpzc{M}$  with its angle $\alpha\in[0,\pi/2)$ since there is a natural  one-to-one correspondence between the elements in $\mathpzc{M}$ and angles in the interval $[0,\pi/2)$.

Time is slotted for $t=0,1,\dots$. The state space  of the qubits  $\mathpzc{S}\subseteq \mathcal{H}$ contains  all possible outcomes of the measurements:
$$
\mathpzc{S}=\Big\{ \cos \alpha|0\rangle +\sin \alpha |1\rangle: \alpha\in[0,\pi) \Big\}.
$$
The state of the qubit held by node $i$ (or simply, qubit $i$) at time $t$ is denoted by $\mathbf{x}_i(t)\in \mathpzc{S}$. Similarly, noting that there is a one-to-one correspondence between a state in $\mathpzc{S}$ and an angle in $[0,\pi)$, we will identify $\mathbf{x}_i(t)$ with its angle whenever convenient.  The network of nodes is interconnected by classical communications. At each time $t$,  node $i$ performs a measurement, denoted $\mathbf{u}_i(t)$ and selected in the set $\mathpzc{M}$, whose outcomes can be exchanged via the classical communication links. The goal is to design efficient rules for the selection of the $\mathbf{u}_i(t)$, so that the $\mathbf{x}_i(t)$ will tend to a common state.

\begin{figure}
\centering
\includegraphics[width=3.6in]{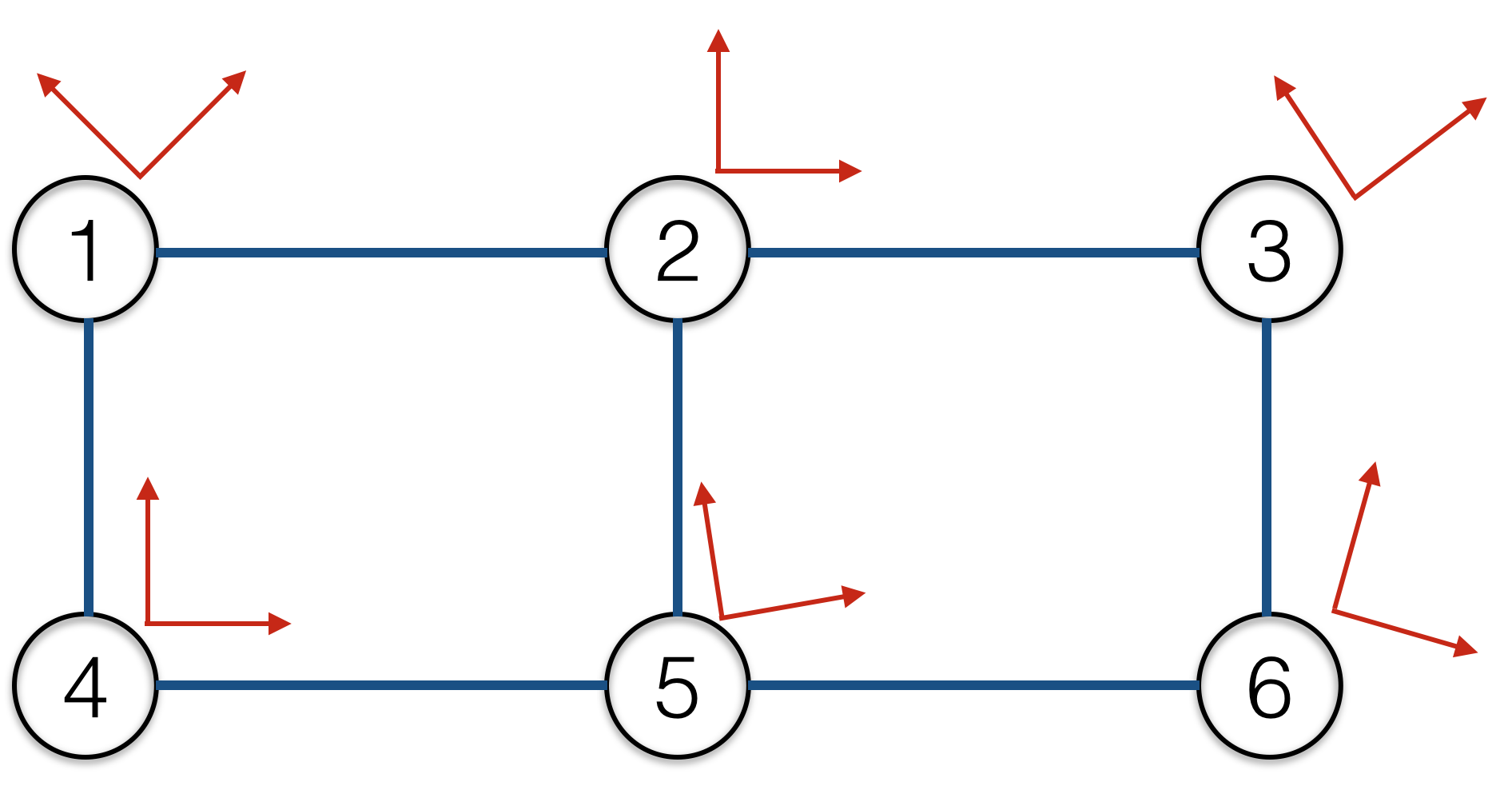}
\caption{An illustration of a six-node quantum hybrid network: There is a qubit at each node, respectively; Projective measurements are performed at the individual qubits;  Nodes are interconnected by classical communication links so that the outcomes of the measurements can be exchanged. }
\label{fig:fig1}
\end{figure}

We give an example of the considered hybrid quantum network with $6$ nodes in Figure \ref{fig:fig1}.

\section{Centralized Solution}\label{sec3}
In this section, we investigate the scenario when the nodes are equipped with all-to-all classical communications and derive the optimal rules for measurement sequence selections at the qubits.

\subsection{Finite Horizon}
We stack the states of the qubits into an $N$ dimensional column vector by $\mathbf{x}(t)=(\mathbf{x}_1(t)\ \dots\ \mathbf{x}_N(t))^\top$. The vector $\mathbf{u}(t)=(\mathbf{u}_1(t)\ \dots\ \mathbf{u}_N(t))^\top$ denotes the selection of measurements performed.  The outcome of the measurement $u_i(t)$ is  $\mathbf{y}_i(t)\in\{\ltimes,\rtimes\}$. We also denote $\mathbf{y}(t)=(\mathbf{y}_1(t)\ \dots\ \mathbf{y}_N(t))^\top$. Suppose the process ends at $t=T$ for some integer $T\geq 1$. The measurement selection decision is denoted by
$$
\sigma=\sigma_0 \times \sigma_1 \dots \times \sigma_{T-1}
$$
where the $\sigma_s$  assigning the value of $\mathbf{u}(s)$. The decision $\sigma_s$ can depend on all information  available by the time slot $s\in\{0,1,\dots,T-1\}$: $\mathbf{x}(t), \mathbf{y}(t)$ for $t=0,\dots,s$ and $\mathbf{u}(t)$ for $t=0,\dots,s-1$. Formally we have
$$
\mathbf{u}(s)=\sigma_s\Big(\mathbf{x}(t), \mathbf{y}(t), t=0,\dots,s; \ \mathbf{u}(t), t=0,\dots,s-1\Big)
$$
with $\sigma_s(\cdot)$ can be an arbitrary function that takes values in $\mathpzc{M}^N$ for $s=0,1,\dots,T-1$.
All such decisions $\sigma$ are put in a set $\Upsilon_T$.

For any fixed measurement decision $\sigma$, the agreement displacement at time $T$ is characterized by the expected network fidelity:
$$
f_\sigma(T)=\mathbb{E}_\sigma\sum_{i,j=1}^N \Big| \big\langle \mathbf{x}_i(T) \big| \mathbf{x}_j(T) \big\rangle\Big|
$$
where $\mathbb{E}_\sigma$ captures all randomness generated by the quantum measurements as well as possible random measurement decisions. The evolution of node states is governed by the quantum measurement principles and can be written as
\begin{align*}
\mathbb{P}\Big(\mathbf{y}(t)=(y_1\ \dots\ y_N)^\top\big| \mathbf{x}(t), \mathbf{u}(t)\Big) = \prod_i\big| \langle\mathbf{x}_i(t) \big|\mathbf{u}_i^{y_i}(t)\rangle\big|^2
\end{align*}
where  $\mathbf{u}_i^{y_i}(t)=\cos (\mathbf{u}_i(t))|0\rangle +\sin (\mathbf{u}_i(t))|1\rangle$ for $y_i=\ltimes$ and $\mathbf{u}_i^{y_i}(t)=\cos (\mathbf{u}_i(t)+\pi/2)|0\rangle +\sin (\mathbf{u}_i(t)+\pi/2)|1\rangle$ for $y_i=\rtimes$. By plain calculation we can further write
\begin{align*}
\mathbb{P}\Big(\mathbf{y}(t)=(y_1\ \dots\ y_N)^\top\big| \mathbf{x}(t), \mathbf{u}(t)\Big)
 = \prod_i  \cos^2 \big( \mathbf{u}_i^{y_i}(t) -\mathbf{x}_i(t)\big).
\end{align*}
Here we have identified $\mathbf{u}_i^{y_i}(t) $ and $\mathbf{x}_i(t)$ with their angles. Note that, the value and distribution of $\mathbf{x}(t+1)$ is fully determined by $\mathbf{y}(t)$ and $\mathbf{u}(t)$. This is to say $\mathbf{x}(t)$ is Markovian.  Finding the policy $\sigma$ that miximizes $f_\sigma(T)$ is a stochastic optimal control problem \cite{bertsekas1978stochastic}.

The  optimal policy  $\sigma^\ast$ that maximizes $f_\sigma(T)$ can be obtained as follows. Clearly $\sigma^\ast$ is Markovian in the sense that
$\mathbf{u}(t)$ depends only on  $\mathbf{x}(t)$ for all $t=0,\dots,T-1$ in the decision profile $\sigma^\ast$. Introduce $u=(u_1\ \dots\ u_N)^\top \in \mathpzc{S}^N$ and $y=(y_1\ \dots\ y_N)^\top \in \{\ltimes, \rtimes\}^N$.  Define a function $\mathbf{Q}(u,y)=(q_1\ \dots\ q_N)^\top\in \mathpzc{S}^N$  by $q_i=u_i$ if $y_i=\ltimes$ and $q_i=u_i+\pi/2 $ if $y_i=\rtimes$. Introduce the cost-to-go function $C(\cdot,\cdot): \mathpzc{S}^N \times \{0,1,\dots,T\} \to \mathbb{R}$ defined by
\begin{align*}C(x,t)= \max_{\sigma\in\Upsilon_T} \mathbb{E}_\sigma \big(\sum_{i,j=1}^N \big| \big\langle \mathbf{x}_i(T) \big| \mathbf{x}_j(T) \big\rangle\big| \big| \mathbf{x}(t)=x \big).
\end{align*}
Then  by a standard dynamic programming argument there holds
\begin{align}\label{1}
C(x,t)=\max_{u\in\mathpzc{M}^N} \sum_{y \in \{\ltimes, \rtimes\}^N} \mathbb{P}\Big(\mathbf{y}(t)=y \big|\mathbf{x}(t)=x, \mathbf{u}(t)=u\Big)  \cdot C(\mathbf{Q}(u,y), t+1)
\end{align}
for $x\in \mathpzc{S}^N$ and $t=0,1,\dots,T-1$. The boundary condition of  (\ref{1}) is
$$
C(x,T)=\sum_{i,j=1}^N \Big| \big\langle {x}_i \big| {x}_j \big\rangle\Big|, \ x=(x_1,\dots,x_n)^\top \in \mathpzc{S}^N.
$$
 The optimal decision $\sigma^\ast$ is given by
\begin{align}
\sigma^\ast_t(\mathbf{x}(t))=\arg \max_{u\in\mathpzc{M}^N} \sum_{y \in \{\ltimes, \rtimes\}^N} \mathbb{P}\Big(\mathbf{y}(t)=y \big|\mathbf{x}(t), \mathbf{u}(t)=u\Big)
\cdot C(\mathbf{Q}(u,y), t+1)
\end{align}
for $t=0,\dots,T-1$.

\subsection{Infinite Horizon}
Next, we consider an infinite horizon  scenario when the  optimality criteria is given by the minimal steps in expectation required for reaching a perfect agreement  in the network. Let $$
\sigma=\sigma_0\times \sigma_1 \times \dots
$$
be a measurement selection policy for the entire time horizon, where for any $s=0,1,\dots$, $\sigma_s$ maps to $\mathpzc{M}^N$ from all available information up to time $s$. All such decisions are put in the set $\Upsilon_\infty$.  Consider the expected number of steps of reaching agreement at the qubits:
$$
g_\sigma=\mathbb{E}_\sigma \Big(\inf_{t} \Big\{t\geq 0:\ \mathbf{x}_1(t)=\dots=\mathbf{x}_N(t) \Big\}\Big).
$$
Clearly there exist simple policies in $\Upsilon_\infty$  under which $g_\sigma$ will be a  finite number. We are interested in the optimal one that minimizes $g_\sigma$.

Recall the definition of $\mathbf{Q}(u,y)$. Similarly, the optimal policy $\sigma^\ast$ that minimizes $g_\sigma$  is Markovian. In fact, it is also   stationary in the sense that $\sigma^\ast_t(\mathbf{x}(t)=x)=\sigma^\ast_s(\mathbf{x}(s)=x)$ for all $s,t\geq 0$.  Define cost-to-go function
\begin{align*}
G(x):=\min_{\sigma\in  \Upsilon_\infty} \mathbb{E}_\sigma
 \Big(\inf_{s} \Big\{s\geq 0:\ \mathbf{x}_1(s+t)=\dots=\mathbf{x}_N(s+t) \Big\} \Big| \mathbf{x}(t)=x\Big).
\end{align*} In this case  the function  $G(x)$ satisfies the following equation \cite{Bertsekas1991}
\begin{align}\label{2}
 G(x)=1+\min_{u\in\mathpzc{M}^N}
 \sum_{y \in \{\ltimes, \rtimes\}^N} \mathbb{P}\Big(\mathbf{y}(t)=y \big|\mathbf{x}(t)=x, \mathbf{u}(t)=u\Big) G(\mathbf{Q}(u,y)).
\end{align}
The optimal  decision $\sigma^\ast$ is given by
\begin{align}
&\sigma^\ast_t(\mathbf{x}(t)=x)=\arg \min_{u\in\mathpzc{M}^N}\sum_{y \in \{\ltimes, \rtimes\}^N} \mathbb{P}\Big(\mathbf{y}(t)=y \big|\mathbf{x}(t)=x, \mathbf{u}(t)=u\Big) G(\mathbf{Q}(u,y)).
\end{align}

\subsection{Computation/Communication  Complexities}
We would like to point out that the derived  optimal network-level rules are conceptually equivalent to the single qubit framework presented in \cite{Shi-PRA14}. Although the centralized   optimal solutions are clear in theory for both finite and infinite time horizons, it is important to understand the  amount of computation and communication  resources required for implementing  them in practice for a considerably large network.

The Bellman equations  (\ref{1}) and (\ref{2})  involve a continuous action set $\mathpzc{M}^N$. Usually this is approximated by a  proper discretization of $\mathpzc{M}$ into a finite set.
For example, we can  let the measurements be selected from \cite{Rabitz-PRA}
\begin{align}
\mathsf{M}_\alpha: \alpha =\frac{j\pi}{2K},\ j=0,\dots,K-1.
\end{align}
This of course means that the resulting policy becomes potentially suboptimal due to smaller action space.  However it is reasonable to believe that a large $K$ would produce policies that can approximate the optimal solution.

Suppose $\mathpzc{M}$ has been discretized into a finite set with $K$ elements. Note that this means that the state space $\mathpzc{S}$ for each qubit is also discretized with $2K$ elements.  We now discuss the finite horizon case in detail. From the computational side, solving the Bellman equation (\ref{1}) relies on recursively along the equation (\ref{1}) computing $C(x,t)$ (and therefore obtain the optimal $\sigma_t^\ast(x)$) from $C(x,t+1)$ for all $t=T-1,\dots,0$ and all $x\in \mathpzc{S}^N$, starting with  the boundary condition
$$
C(x,T)=\sum_{i,j=1}^N \big| \big\langle {x}_i \big| {x}_j \big\rangle\big|=\sum_{i,j=1}^N \big|\cos(x_i-x_j) \big|^2.
$$
The number of algebraic operations required in such process inevitably grows faster than $K^N$. Therefore, practically it is almost  impossible to numerically solve  the Bellman equation (\ref{1}) and obtain the optimal policy for a large network. In fact, even if the computation can be done off line, preserving the optimal policy relies on $O((2K)^N  T \log K)$ bits of memory. Since each node relies on the states of all other nodes to carry out the optimal policy,  the network requires all-to-all communications with  $O\big(N^2)$ bits of transmissions per step.

\section{Distributed Solution}\label{sec4}
In this section, we discuss distributed solutions to the considered qubit consensus problem  in the sense that nodes communicate  with a few neighbours locally  and then make measurement selection decisions individually.

\subsection{The Algorithm}
We assume that
there is a connected underlying graph $\mathrm{G}=(\mathrm{V}, \mathrm{E})$ with node set $\mathrm{V}$ and edge set $\mathrm{E}$ representing  the classical communication links among the nodes, where a link $\{i,j\}\in\mathrm{E}$ specifies that nodes $i$ and $j$ can exchange information their states. We denote $\mathrm{N}_i:=\{j: \{i,j\}\in\mathrm{E}\}$ as the neighbour set of node $i$. We also define
\begin{equation*}
a \mod \pi/2 =
  \begin{cases}
   a       & \quad \text{if }   a\in [0,\pi/2),\\
   a- \pi/2 & \quad \text{if }   a\in [\pi/2,\pi)\\
  \end{cases}
\end{equation*}
and
\begin{equation*}
a \mod \pi =
  \begin{cases}
   a       & \quad \text{if }   a\in [0,\pi),\\
   a- \pi & \quad \text{if }   a\in [\pi,2\pi).\\
  \end{cases}
\end{equation*}

We propose the following algorithm.

\medskip

\noindent {\bf Pairwise Qubit Projection (PQP)}. (i) At each $t$, a node $i$ is drawn uniformly at random from the set $\mathrm{V}$, and then node $j$ is selected uniformly at random from the set $\mathrm{N}_i$; (ii) The selected pair of nodes $i$ and $j$ exchanges their current states $\mathbf{x}_i(t)$ and $\mathbf{x}_j(t)$; (iii) Nodes $i$ and $j$ apply projective measurements $$
\mathbf{u}_i(t)=\mathbf{u}_j(t)=(\mathbf{x}_i(t)+\mathbf{x}_j(t))/2 \mod \pi/2
$$ and all other nodes keep their current states.

\medskip

We remark that the above algorithm is clearly inspired by the class of gossiping algorithms for classical communication networks and open quantum  networks \cite{Boyd2006,Ticozzi,Shi-TON2016}. This proposed  algorithm can be realized in fully distributed manner in the sense that nodes even need not to share a common clock and can simply follow independent Poisson processes to wake up \cite{Boyd2006}. Moreover, the pair section process can also be made deterministic and multiple disjoint  pairs can be selected at a given time, which will not change the nature of the algorithm and actually can speed up the algorithm.  The involved projective measurements  introduce new type of randomness in the algorithm, which makes the PQP algorithm differ from the previous algorithms \cite{Boyd2006,Ticozzi,Shi-TON2016} at a fundamental level.

\subsection{State Evolution}
Let $\mathbf{x}(t)$ be driven by the proposed PQP algorithm. Suppose node  pair $\{i,j\}$ is selected at time $t$. Then from the quantum measurement postulate,  independently among $m\in \{i,j\}$ we have
\begin{equation*}
\mathbf{x}_m(t+1)=\frac{\mathbf{x}_i(t)+\mathbf{x}_j(t)}{2}
\end{equation*}
with probability $\cos ^2\frac{\mathbf{x}_i(t)-\mathbf{x}_j(t)}{2}$, and
\begin{equation*}
\mathbf{x}_m(t+1)=\frac{\mathbf{x}_i(t)+\mathbf{x}_j(t)+\pi}{2} \mod \pi
\end{equation*}
with probability  $\sin ^2\frac{\mathbf{x}_i(t)-\mathbf{x}_j(t)}{2}$. The following result holds.

\begin{theorem}\label{theorem-almostsure}
Using the PQP algorithm, the hybrid quantum network reaches an agreement almost surely  in the sense that
$$
 \mathbb{P}\Big( \lim_{t\to \infty}|\mathbf{x}_i(t)-\mathbf{x}_j(t)|=0 \Big)=1
$$
for all $i,j\in\mathrm{V}$.
\end{theorem}

We defer the proof of Theorem \ref{theorem-almostsure} to the end of this section.

\begin{figure}
\centering
\includegraphics[width=3.9in]{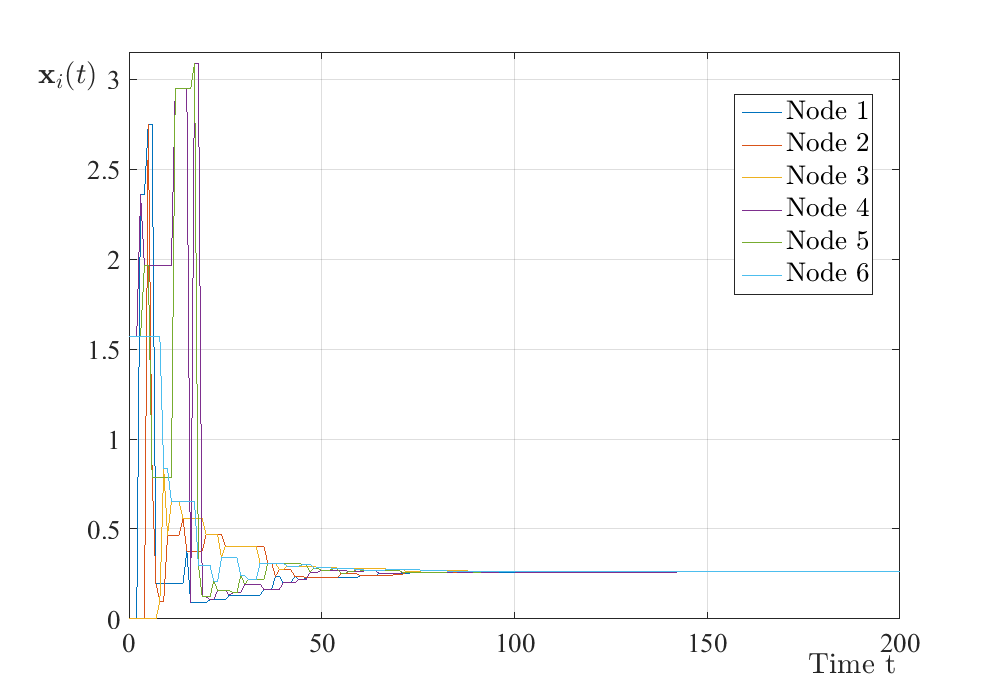}
\caption{ A sample path of $\mathbf{x}_i(t), i=1,\dots,6$ along the PQP algorithm with initial value $\mathbf{x}_1(0)=\mathbf{x}_2(0)=\mathbf{x}_3(0)=0$ and $\mathbf{x}_1(0)=\mathbf{x}_2(0)=\mathbf{x}_3(0)=\pi/2$. }
\label{fig:fig2}
\end{figure}

\begin{figure}
\centering
\includegraphics[width=3.9in]{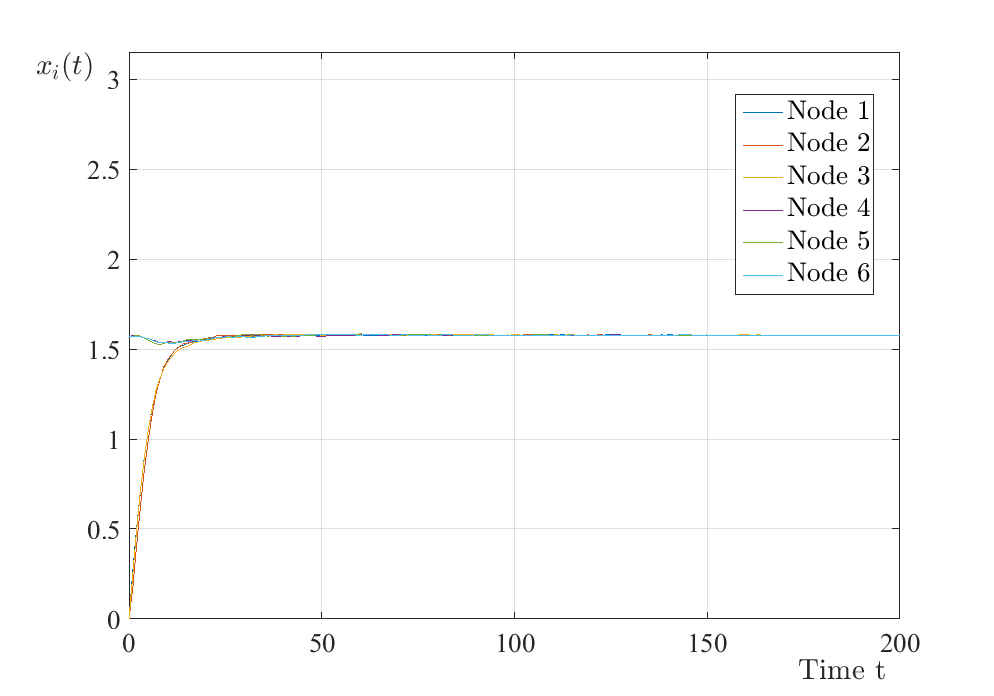}
\caption{Trajectories of  $x_i(t):=\mathbb{E}(\mathbf{x}_i(t)), i=1,\dots,6$ along the PQP algorithm with initial value $\mathbf{x}_1(0)=\mathbf{x}_2(0)=\mathbf{x}_3(0)=0$ and $\mathbf{x}_1(0)=\mathbf{x}_2(0)=\mathbf{x}_3(0)=\pi/2$. }
\label{fig:fig3}
\end{figure}

For the hybrid quantum network illustrated in Figure \ref{fig:fig1}, we plot a sample path at which consensus is reached and the trajectories of the expected states of the qubits, respectively, in Figure \ref{fig:fig2} and Figure \ref{fig:fig3}.

%
%

\subsection{Density Evolution}
Now we introduce $\pmb{\rho}_i(t)$ as the density operator corresponding to $\mathbf{x}_i(t)\in\mathpzc{S}$. Viewing also $\mathbf{x}_i(t)$ as its angle in $[0,\pi)$, we can formally write:
\begin{equation}\label{11}
\pmb{\rho}_i(t)=\begin{bmatrix}
\cos^2 \mathbf{x}_i(t)& \cos \mathbf{x}_i(t)\sin\mathbf{x}_i(t)\\
\cos \mathbf{x}_i(t)\sin\mathbf{x}_i(t) & \sin^2\mathbf{x}_i(t)\end{bmatrix}
\end{equation}
We also define $\rho_i(t)=\mathbb{E}\big\{\pmb{\rho}_i(t) \big\}$, where $\mathbb{E}$  is subject to the classical measure $\mathbb{P}$ capturing all randomness in the node pair selection process and in the quantum projective measurements. Note that  $\pmb{\rho}_i(t)$ is always a pure state. All the outcomes of the projective measurements have to be read out for carrying out the algorithm. Nonetheless $\rho_i(t)$ describes the distribution of   $\pmb{\rho}_i(t)$ under the measure $\mathbb{P}$. We stack $\pmb{\rho}(t)=(\pmb{\rho}_1(t)\ \dots \ \pmb{\rho}_N(t))^\top$
and ${\rho}(t)=({\rho}_1(t)\ \dots \ {\rho}_N(t))^\top$ as vectors of $2\times 2$ density operators.

It turned out that it is more convenient to investigate the evolution of the $\mathbf{x}(t)$ from the corresponding density operators, whose original update is in fact rather complex. Let $L_{\mathrm{G}}$ be the  Laplacian of the graph $\mathrm{G}$,  defined by $[L_{\mathrm{G}}]_{ij}=-(1/|\mathrm{N}_i|+1/|\mathrm{N}_j|)$ for $\{i,j\}\in\mathrm{E}$, $[L_{\mathrm{G}}]_{ij}=0$ for $i\neq j$ with $\{i,j\}\notin \mathrm{E}$, and $[L_{\mathrm{G}}]_{ii}=\sum_{j=1}^N [L_{\mathrm{G}}]_{ij}$. We have the following result.

\begin{figure}
\centering
\includegraphics[width=3.9in]{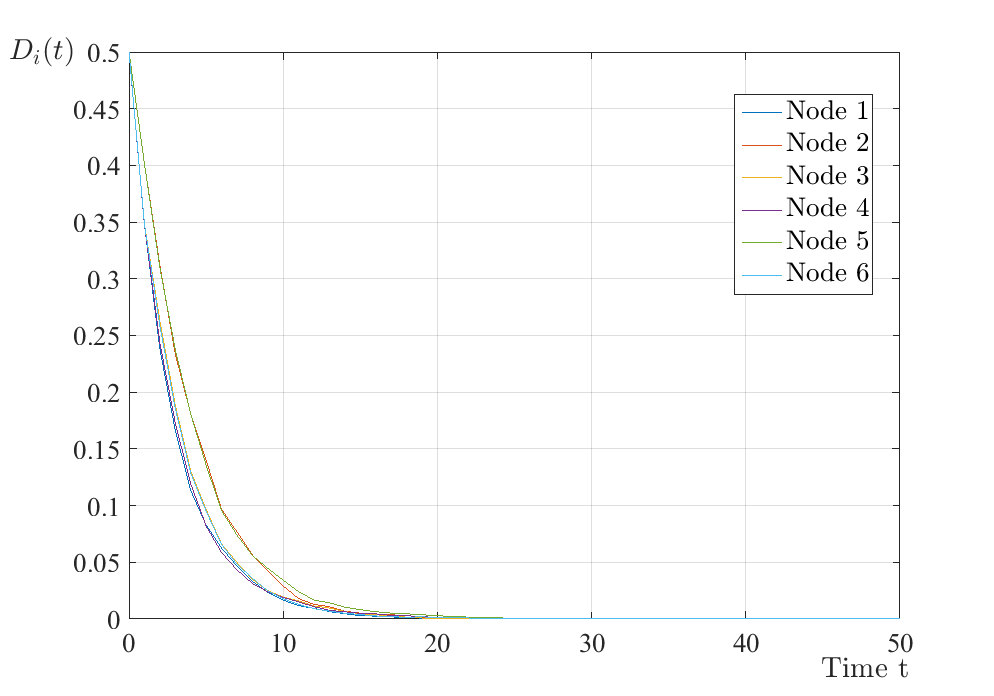}
\caption{Trajectories of  $D_i(t), i=1,\dots,6$ along the PQP algorithm with initial value $\mathbf{x}_1(0)=\mathbf{x}_2(0)=\mathbf{x}_3(0)=0$ and $\mathbf{x}_1(0)=\mathbf{x}_2(0)=\mathbf{x}_3(0)=\pi/2$. }
\label{fig:fig4}
\end{figure}

\begin{theorem}\label{theorem-operator}
The density vector sequence $\big({\rho}(t)\big)_{t\geq 0}$ satisfies
\begin{align*}
{\rho}(t+1)=\big(I_N -L_{\mathrm{G}} \big)\otimes I_2 {\rho(t)},\ t\geq 0.
\end{align*}
Consequently, we have
\begin{align*}
\lim_{t\to \infty} \rho_i(t)=\frac{\sum_{i=1}^N \rho_i(0)}{N}, \ i\in\mathrm{V}
\end{align*}
with an exponential rate at $1-\lambda_2(L_{\mathrm{G}})$, where $\lambda_2(L_{\mathrm{G}})$ is the smallest positive eigenvalue of $L_{\mathrm{G}}$.
\end{theorem}
{\it Proof.} Suppose node pair $\{i,j\}$ is selected at time $t$. Then based on  (\ref{11}), we obtain

\begin{align}\label{16}
\mathbb{E}\big\{\pmb{\rho}_i(t+1) \big| \pmb{\rho}(t)\big\}
&=
\mathbb{E}\big\{\pmb{\rho}_j(t+1) \big| \pmb{\rho}(t)\big\}\nonumber\\
&=\cos ^2\frac{\mathbf{x}_i(t)-\mathbf{x}_j(t)}{2}  \begin{bmatrix}
\cos^2 \frac{\mathbf{x}_i(t)+\mathbf{x}_j(t)}{2}& \cos \frac{\mathbf{x}_i(t)+\mathbf{x}_j(t)}{2}\sin\frac{\mathbf{x}_i(t)+\mathbf{x}_j(t)}{2}\\
\cos \frac{\mathbf{x}_i(t)+\mathbf{x}_j(t)}{2}\sin\frac{\mathbf{x}_i(t)+\mathbf{x}_j(t)}{2} & \sin^2\frac{\mathbf{x}_i(t)+\mathbf{x}_j(t)}{2}\end{bmatrix}\nonumber\\
&+\sin ^2\frac{\mathbf{x}_i(t)-\mathbf{x}_j(t)}{2} \begin{bmatrix}
\sin^2 \frac{\mathbf{x}_i(t)+\mathbf{x}_j(t)}{2}& -\cos \frac{\mathbf{x}_i(t)+\mathbf{x}_j(t)}{2}\sin\frac{\mathbf{x}_i(t)+\mathbf{x}_j(t)}{2}\\
-\cos \frac{\mathbf{x}_i(t)+\mathbf{x}_j(t)}{2}\sin\frac{\mathbf{x}_i(t)+\mathbf{x}_j(t)}{2} & \cos^2\frac{\mathbf{x}_i(t)+\mathbf{x}_j(t)}{2}\end{bmatrix}\nonumber\\
&=\frac{1}{2}\begin{bmatrix}
\cos^2 \mathbf{x}_i(t)& \cos \mathbf{x}_i(t)\sin\mathbf{x}_i(t)\\
\cos \mathbf{x}_i(t)\sin\mathbf{x}_i(t) & \sin^2\mathbf{x}_i(t)\end{bmatrix}
+\frac{1}{2}\begin{bmatrix}
\cos^2 \mathbf{x}_j(t)& \cos \mathbf{x}_j(t)\sin\mathbf{x}_j(t)\\
\cos \mathbf{x}_j(t)\sin\mathbf{x}_j(t) & \sin^2\mathbf{x}_j(t)\end{bmatrix}
\nonumber\\
&= \frac{1}{2}\pmb{\rho}_i(t) +\frac{1}{2}\pmb{\rho}_j(t),
\end{align}
where the third equality holds from elementary sum-to-product trigonometric formulas.

We further obtain
 \begin{align*}
{\rho}(t+1)=\big(I_N -L_{\mathrm{G}} \big)\otimes I_2 {\rho(t)}.
\end{align*}
by  collecting all events at the pairs of nodes. The convergence statement aligns with the same argument as used in \cite{Boyd2006}. This concludes the proof. \hfill$\square$

 Theorem \ref{theorem-operator} shows that in the operator space,  $\mathbb{E}\{\pmb{\rho}(t)\}$ simply follows a linear time-invariant system and eventually leads to an average consensus. This result is related  but also in contrast to the work of \cite{Ticozzi,Shi-TAC}, which showed that a network of qubits interconnected by local environments can be driven to a consensus of their individual reduced states.  The evolution of $\mathbb{E}\{\mathbf{x}(t)\}$ is on the other hand highly complex for which even a nonlinear recursive form  is out of reach.

An illustration of  Theorem \ref{theorem-operator} is presented below for  the hybrid quantum network in Figure \ref{fig:fig4}, where we plot $$
D_i(t):= \Big\| \rho_i(t) - \frac{\sum_{i=1}^N \rho_i(0)}{N}\Big\|_2
$$
for $i=1,\dots,6$, respectively.

\subsection{Proof of Theorem \ref{theorem-almostsure}}

Define  $p_{ij}=(1/|\mathrm{N}_i|+1/|\mathrm{N}_j|)$ for $\{i,j\}\in\mathrm{E}$ as the probability of link $\{i,j\}$ being selected at a given time.  Introduce $\mathbf{h}(t):=\sum_{\{i,j\}: i< j} {\rm Tr} \big(\pmb{\rho}_i(t)\pmb{\rho}_j(t)\big)$. Then we have
\begin{align}\label{19}
&\mathbb{E}\Big( \mathbf{h}(t+1) \Big| \pmb{\rho}(t) \Big)\nonumber\\
&=\sum_{\{k,m\}\in \mathrm{E}} p_{km} \mathbb{E}_{km}\Big( \mathbf{h}(t+1) \Big| \pmb{\rho}(t) \Big)\nonumber\\
&=\sum_{\{k,m\}\in \mathrm{E}} p_{km} \mathbb{E}_{km} \bigg[  {\rm Tr}\big(\pmb{\rho}_k(t+1)\pmb{\rho}_m(t+1)\big)+\sum_{\{i,j\}:i<j, \{i,j\}\neq \{k,m\}} {\rm Tr} \big(\pmb{\rho}_i(t+1)\pmb{\rho}_j(t+1)\big) \bigg| \pmb{\rho}(t) \bigg]\nonumber\\
&=\sum_{\{k,m\}\in \mathrm{E}} p_{km} \Bigg({\rm Tr}\bigg[ \mathbb{E}_{km} \big(\pmb{\rho}_k(t+1)\pmb{\rho}_m(t+1)\big) \bigg| \pmb{\rho}(t) \bigg]+{\rm Tr}\bigg[ \sum_{\{i,j\}: i<j, \{i,j\}\neq \{k,m\}} \mathbb{E}_{km} \big(\pmb{\rho}_i(t+1)\pmb{\rho}_j(t+1)\big) \bigg| \pmb{\rho}(t) \bigg]\Bigg)
\end{align}
where in the first equality the $ \mathbb{E}_{km}$ is subject to the randomness generated by quantum measurements at nodes $k$ and $m$, and in the last equality we have used the fact that trace and expectation commute due to their linearity.

Proceeding with the first of the two trace terms in the right-hand side of (\ref{19}), we have
\begin{align}\label{17}
&{\rm Tr}\bigg[ \mathbb{E}_{km} \big(\pmb{\rho}_k(t+1)\pmb{\rho}_m(t+1)\big) \bigg| \pmb{\rho}(t) \bigg]\nonumber\\
&= {\rm Tr} \bigg[\mathbb{E}_{km}  \big(\pmb{\rho}_k(t+1)\big)  \mathbb{E}_{km}  \big(\pmb{\rho}_m(t+1)\big) \bigg| \pmb{\rho}(t) \bigg] \nonumber\\
 &={\rm Tr} \Big( \frac{1}{2}\pmb{\rho}_k(t) +\frac{1}{2}\pmb{\rho}_m(t)\Big) \Big( \frac{1}{2}\pmb{\rho}_k(t) +\frac{1}{2}\pmb{\rho}_m(t)\Big)\nonumber\\
 &= \frac{1}{2}+\frac{1}{2} {\rm Tr}  \Big(\pmb{\rho}_k(t)\pmb{\rho}_m(t)\Big),
\end{align}
where the first equality is due to independence of the outcome of the quantum measurements at nodes $k$ and $m$, and the second equality utilizes (\ref{16}). Meanwhile, considering the second trace term in the right-hand side of (\ref{19}), it is easy to conclude from  (\ref{16}) that
\begin{align}\label{18}
{\rm Tr}\bigg[ \sum_{\{i,j\}:i<j, \{i,j\}\neq \{k,m\}} \mathbb{E}_{km} \big(\pmb{\rho}_i(t+1)\pmb{\rho}_j(t+1)\big) \bigg| \pmb{\rho}(t) \bigg]=
 \sum_{\{i,j\}:i<j, \{i,j\}\neq \{k,m\}} {\rm Tr}\big(\pmb{\rho}_i(t)\pmb{\rho}_j(t)\big).
\end{align}
As a result, from (\ref{19}), (\ref{17}) and (\ref{18}) we have
\begin{align}\label{20}
\mathbb{E}\Big( \mathbf{h}(t+1) \Big| \pmb{\rho}(t) \Big)&=\sum_{\{k,m\}\in \mathrm{E}} p_{km} \bigg[\mathbf{h}(t) +\frac{1}{2}\Big( 1-{\rm Tr}  \big(\pmb{\rho}_k(t)\pmb{\rho}_m(t)\big) \Big) \bigg]\nonumber\\
 &=\mathbf{h(t)}+\sum_{\{k,m\}\in \mathrm{E}}  \frac{ p_{km}}{2}\Big( 1-{\rm Tr}  \big(\pmb{\rho}_k(t)\pmb{\rho}_m(t)\big) \Big)
\end{align}

Since  ${\rm Tr}  \big(\pmb{\rho}_k(t)\pmb{\rho}_m(t)\big)\leq 1$ always holds, (\ref{20}) implies that $\big\{\mathbf{h}(t)\big\}$ is a submartingale. Moreover, $\mathbb{E}\big(  \mathbf{h}(t)\big)\leq N(N-1)/2$ for all $t$ by the definition of $\mathbf{h}(t)$. By the Martingale Convergence Theorem (Theorem 5.2.8, \cite{Durrett-book}), $\mathbf{h}(t)$ converges to a finite limit almost surely. We can further invoke the Dominated  Convergence Theorem (e.g., Exercise 2.3.7, \cite{Durrett-book}) to yield that, $\mathbb{E}(\mathbf{h}(t))$ converges to a finite limit. Hence, (\ref{20}) implies
\begin{align*}
 \mathbb{E}\Big( \mathbf{h}(t+1) \Big)  =\mathbb{E}\Big(\mathbf{h}(t)\Big)+ \mathbb{E}\bigg[\sum_{\{k,m\}\in \mathrm{E}}  \frac{ p_{km}}{2}\Big( 1-{\rm Tr}  \big(\pmb{\rho}_k(t)\pmb{\rho}_m(t)\big) \Big)\bigg],
\end{align*}
so  $\lim_{t\to \infty}\mathbb{E} \Big({\rm Tr}  \big(\pmb{\rho}_k(t)\pmb{\rho}_m(t)\big) \Big)=1$ for all $\{k,m\}\in \mathrm{E}$. However, as ${\rm Tr}  \big(\pmb{\rho}_k(t)\pmb{\rho}_m(t)\big) \leq 1$ is a sure event, we conclude for any $\epsilon>0$ that
\begin{align}
\lim_{t\to \infty}\mathbb{P} \Big(  {\rm Tr}  \big(\pmb{\rho}_k(t)\pmb{\rho}_m(t)\big) \geq 1- \epsilon \Big)=1,
\end{align}
i.e., ${\rm Tr}  \big(\pmb{\rho}_k(t)\pmb{\rho}_m(t)\big)$ converges to $1$ in probability for all $\{k,m\}\in \mathrm{E}$.

Finally, we notice that   ${\rm Tr}  \big(\pmb{\rho}_k(t)\pmb{\rho}_m(t)\big)=\cos^2(\mathbf{x}_k(t) -\mathbf{x}_m(t))$. Therefore, ${\rm Tr}  \big(\pmb{\rho}_k(t)\pmb{\rho}_m(t)\big)$ converging to one in probability  is equivalent to that $\mathbf{x}_k(t) -\mathbf{x}_m(t)$ converging to zero in probability. While $\mathrm{G}$ is a connected graph, we further know that $\mathbf{x}_i(t) -\mathbf{x}_j(t)$ converges  to zero in probability for all $i,j\in\mathrm{V}$. This immediately implies that $\mathbf{h}(t)$ will converge to $N(N-1)/2$ in probability. However, we have known as a fact that $\mathbf{h}(t)$ converges in the almost sure sense. Therefore, $\mathbf{h}(t)$ must converge to $N(N-1)/2$ almost surely, or equivalently, $\mathbf{x}_i(t) -\mathbf{x}_j(t)$ converging to zero almost surely for all $i,j\in\mathrm{V}$. The desired theorem holds and we have now completed the proof.

\section{Conclusions}\label{sec5}
We have  considered  a consensus seeking problem over a quantum hybrid  network.   A number  of nodes each holding a qubit apply projective measurements  and the measurement results  are exchanged  via classical communications.  Centralized optimal path planning for the network with all-to-all classical communications were derived by  stochastic optimal control approach, whose overwhelming computation and communication complexities were shown for a large network. A distributed  Pairwise Qubit Projection (PQP) algorithm was also proposed along which the qubit states can be  driven to a consensus almost surely along the proposed PQP algorithm. Future work includes generalization of the optimal control and distributed control approaches to hybrid quantum networks in the presence of  quantum links as entangled pairs for improving the efficiency and scalability of such networks in applications.

\end{document}